\documentclass{article}
\usepackage[T1]{fontenc}
\usepackage[utf8]{inputenc}
\usepackage[]{ismir} 
\usepackage{amsmath,cite,url}
\usepackage{graphicx}
\usepackage{color}
\usepackage{pgfplots}
\usepgfplotslibrary{groupplots,statistics}
\usetikzlibrary{calc}
\pgfplotsset{compat=1.18}
\title{MuScriptor: An Open Model for Multi-Instrument Music Transcription}

\multauthor
  {Simon Rouard$^{1,3\dagger}$ \hspace{1cm} Michael Krause$^{2\dagger}$ \hspace{1cm} Axel Roebel$^3$}
  {{\bf Carl-Johann Simon-Gabriel$^2$ \hspace{1cm} Alexandre Défossez$^1$}\\
  $^1$ Kyutai \hspace{1cm} $^2$ Mirelo AI \hspace{1cm} $^3$ UMR STMS, IRCAM-CNRS Sorbonne Univ.\\
  {\tt\small simon@kyutai.org, michael@mirelo.ai}
  }

\def\authorname{S. Rouard, M. Krause, A. Roebel, C.-J. Simon-Gabriel, and A. Défossez}

\usepackage[bookmarks=false,pdfauthor={\authorname},pdfsubject={\pdfsubject},hidelinks]{hyperref}
\usepackage{booktabs}
\usepackage{multirow}
\usepackage{pgfplots}
\pgfplotsset{compat=1.18}
\newcommand{\refsection}[1]{Section~\ref{#1}}
\newcommand{\reftable}[1]{Table~\ref{#1}}
\newcommand{\reffigure}[1]{Figure~\ref{#1}}

\newcommand{\dsynth}{$\mathcal{D}_{\text{Synth}}$}
\newcommand{\dreal}{$\mathcal{D}_{\text{Real}}$}
\newcommand{\drl}{$\mathcal{D}_{\text{RL}}$}
\newcommand{\dtest}{$\mathcal{D}_{\text{Test}}$}
\newcommand{\acfg}{\alpha_{\text{CFG}}}

\begin{document}

\maketitle

\begin{abstract}
Existing methods for automatic music transcription are often limited to single-instrument recordings or fail on complex, real music mixes. Although previous work utilizes synthetic training data, the resulting models generalize poorly, leading to largely unusable transcription output in realistic, multi-instrument settings.
In this work, we analyze the effectiveness of synthetic data for pre-training while combining it with fine-tuning on real music audio and post-training using reinforcement learning. We further introduce conditioning on instrument presence to customize transcriptions. Finally, we release {MuScriptor}, an open-weight multi-instrument music transcription model that works on real-world music recordings from across a diverse range of musical genres.
\end{abstract}

\vspace{-0.15cm}
\section{Introduction}
\vspace{-0.1cm}

The task of Automatic Music Transcription (AMT) consists of converting an audio recording of a piece of music into some kind of symbolic representation, typically MIDI. While significant progress has been made in transcribing single-instrument recordings (specifically for piano \cite{maestro}, guitar \cite{guitarset}, and drums \cite{unsupervised_drums, drums2}), general-purpose transcription for multiple instruments remains a significant challenge. Transcribing multi-instrument music from diverse musical genres requires models to handle a vast range of timbres, overlapping frequencies, and audio effects (such as distortion on electric guitars), across a wide sonic spectrum.

A primary bottleneck in building multi-instrument AMT systems is the scarcity of music audio with aligned note annotations. Recent works such as MT3 \cite{mt3} attempt to solve this by combining small real-world datasets with large-scale synthetic data. For example, MT3 utilizes approximately 1500 hours of synthetic data combined with only 250 hours of (mostly single-instrument) real-world recordings. Although these models perform well on synthetic test sets, their performance often degrades significantly when applied to real-world audio. This suggests a critical domain shift between synthesized MIDI and the complexities of professional music productions.

As a consequence, existing models for multi-instrument music transcription are usually too error-prone to be used for downstream applications. However, a general-purpose music transcription model could be an essential tool for musicians and musicologists, as well as enable new work on generative modeling and various music information retrieval tasks like chord or key recognition.

\begin{figure*}[htbp]
    \centering
    \includegraphics[width=0.95\textwidth]{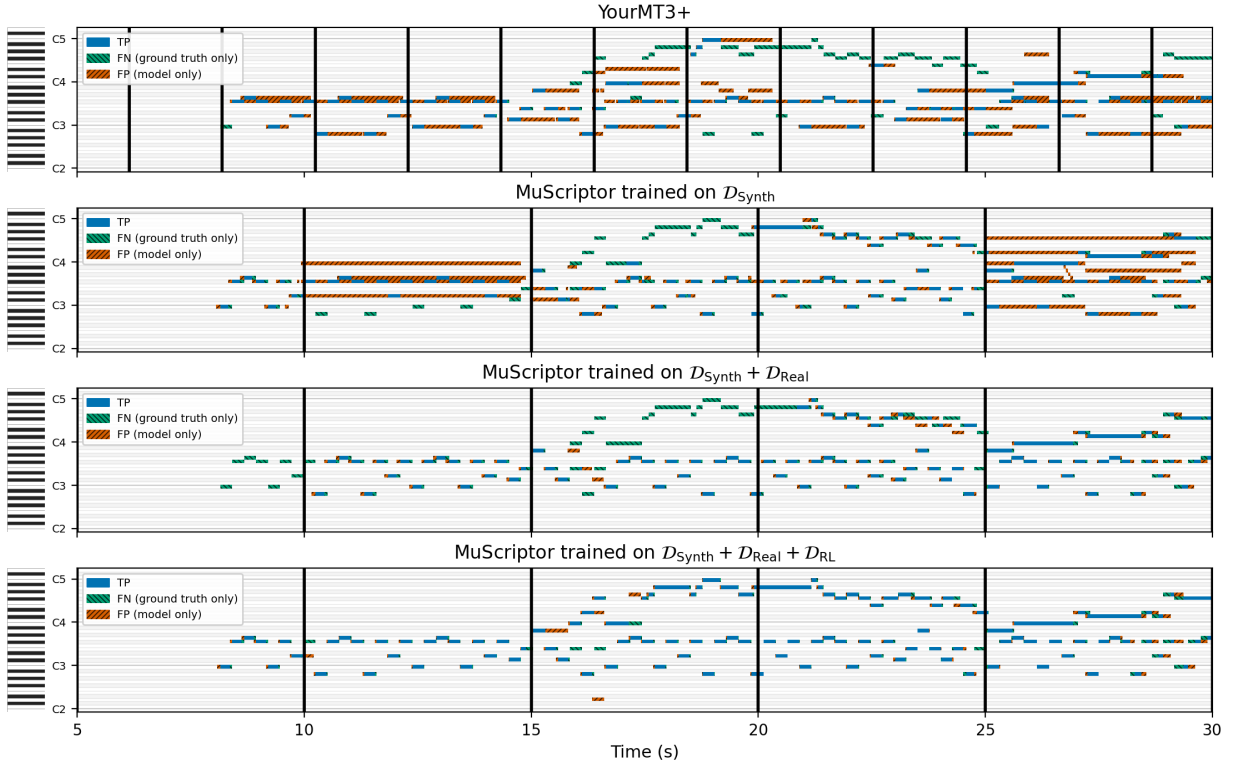}
    \vspace{-0.5cm}
    \caption{Piano roll for the guitar notes in a song from our test set \dtest. Blue indicates true positives (a note was correctly detected by the model), green indicates false negatives (the model has missed a note), and red indicates false positives (the model predicted a wrong note). Vertical lines indicate the segments on which the model is evaluated.}
    \label{fig:pianoroll_song20}
\end{figure*}

In this paper, we investigate the effectiveness of training on synthetic data for music transcription and show that it can be very useful for pre-training, but not sufficient to enable general-purpose transcription. We further collect a large dataset of 170k real music recordings from a wide variety of genres (from classical to heavy metal), featuring audio and aligned note annotations. Through this, we are able to train an effective multi-instrument transcription model, which we release to the public. \reffigure{fig:pianoroll_song20} gives a qualitative impression of the improvements we achieve compared to state-of-the-art AMT models.{\renewcommand{\thefootnote}{$\dagger$}\footnotetext{Equal contribution}}

We forego complex architectural tweaks and instead opt for a simple yet effective decoder-only transformer architecture. We compare two primary training regimes: training exclusively on real data and pre-training on synthetic data followed by real data fine-tuning. Furthermore, we post-train our models with reinforcement learning on a curated subset of 300 high quality transcribed pieces to improve results. Finally, we enable conditioning on instrument presence, allowing users to customize transcription.

\noindent Overall, we make the following contributions:

\textbf{1. An Open-Weight Transcription Model:} We release {MuScriptor}, a model (including weights and inference code)\footnote{\url{https://github.com/muscriptor/muscriptor}} for general-purpose multi-instrument music transcription across diverse musical styles. To train it, we gather a dataset of 170k recordings (11k hours) with aligned note annotations that covers most musical genres, alongside a synthetic dataset of 1.45M MIDIs.

\textbf{2. Analysis of Synthetic Pre-training:} We provide a comparative study on the benefits and limitations of pre-training on synthetic data versus training on real-world recordings with aligned note annotations.

\textbf{3. Alignment via Reinforcement Learning:} We apply a GRPO-like algorithm to our model to align it with a small dataset of high-quality transcriptions.

\textbf{4. Instrument-conditioned Transcription:} By optionally conditioning on instrument presence, we stabilize predictions across different segments of a music recording and allow users to customize their transcription output.

\section{Related Work}

AMT models historically evolved from highly specialized instrument specific architectures towards generalized sequence modeling paradigms. Early research focused on single-instrument scenarios (most notably solo piano) using traditional methods such as hidden markov models \cite{hmm_piano}, non negative matrix factorization \cite{nmf_piano}, or support vector machines \cite{svm_piano}. With the rising prominence of deep learning, AMT approaches began utilizing recurrent neural networks \cite{rnn_piano} or convolutional neural networks to classify the presence of pitches within a specific audio frame \cite{cnn_piano}. One example is the Onsets and Frames architecture \cite{onsets_and_frames}, which demonstrated that treating note start times and sustain durations as separate but related objectives significantly improved accuracy. However, these models relied on a piano roll representation of transcriptions, which becomes computationally expensive and sparse when scaled to multiple instruments and long durations.

An alternative paradigm emerged with the introduction of sequence to sequence modeling for music. Instead of classifying every time frequency bin, models began to treat transcription as a language modeling task by predicting a series of discrete MIDI-like events, e.g., for piano notes \cite{s2s_piano}. For multi-instrument transcription, the Multi-Task Multitrack Music Transcription (MT3) framework \cite{mt3} utilizes an encoder-decoder Transformer to ingest audio spectrograms and autoregressively output a discrete stream of MIDI-like tokens representing pitch, timing, and instrument class. MT3 is trained on a mixture of instrument-specific data as well as multi-instrument synthesized data. Although MT3 established a new baseline, its reliance on synthetic data mixtures causes deteriorating results on real data. In practice, most outputs are not accurate enough to be useful for downstream tasks. We show in this paper that training an MT3-like model on real multi-instrument data greatly improves the quality of the transcriptions. 

Following MT3, recent work introduced improvements in the sequence-to-sequence framework. YourMT3+ \cite{ymt3_plus} improves the MT3 baseline by replacing the standard encoder with a hierarchical attention transformer operating in the time-frequency domain, combined with a Mixture of Experts (MoE) layer and cross-dataset stem augmentation. Nevertheless, the authors of YourMT3+ still encountered poor performance on real-world datasets. To date, MT3 is still a strong baseline that few approaches managed to beat in the 2025 AMT Challenge \cite{amt_challenge_2025}. For instance, MIROS extends the YourMT3+ framework by replacing the spectrogram encoder with MusicFM \cite{musicfm}, a self-supervised audio foundation model. 

\begin{figure*}[t] 
\centering
\begin{tikzpicture}
\begin{axis}[
    ybar, 
    width=\textwidth, 
    height=5cm, 
    bar width=5pt, 
    ylabel={$\%$ of tracks},
    ymin=0, ymax=65,
    xtick={1,2,3,4,5,6,7,8,9,10,11,12,13,14,15,16,17,18,19,20,21,22,23,24,25,26,27,28,29,30,31,32,33},
    xticklabels={Acoustic Piano, Distorted E-Guitar, E-Bass, Drums, Acoustic Guitar, Clean E-Guitar, Voice, String Ensemble, Flute, Organ, Synth Lead, Acoustic Bass, Trumpet, Chromatic Percussion, Soprano/Alto Sax, Violin, Synth Pad, E-Piano, Trombone, Clarinet, French Horn, Tenor Sax, Cello, Oboe, Brass Ensemble, Synth Strings, Bassoon, Viola, Tuba, Baritone Sax, Timpani, Contrabass, Harp},
    enlarge x limits=0.02,
    nodes near coords,
    nodes near coords style={
        font=\tiny, 
        rotate=90, 
        anchor=west, 
        /pgf/number format/.cd, fixed, precision=3 
    },
    tick label style={font=\tiny, rotate=30, anchor=east},
    xticklabel style={font=\scriptsize, rotate=15, anchor=north east},
    label style={font=\scriptsize},
    every axis plot/.append style={fill=blue!40, draw=blue!60},
]
\addplot coordinates {
    (1, 51.03)
    (2, 46.75)
    (3, 42.80)
    (4, 31.62)
    (5, 24.33)
    (6, 21.81)
    (7, 21.09)
    (8, 16.05)
    (9, 15.05)
    (10, 09.22)
    (11, 07.78)
    (12, 07.43)
    (13, 06.83)
    (14, 06.22)
    (15, 05.80)
    (16, 05.76)
    (17, 05.68)
    (18, 05.19)
    (19, 04.86)
    (20, 04.56)
    (21, 04.11)
    (22, 04.02)
    (23, 03.92)
    (24, 03.69)
    (25, 03.30)
    (26, 03.27)
    (27, 02.70)
    (28, 02.66)
    (29, 02.63)
    (30, 02.19)
    (31, 02.03)
    (32, 01.74)
    (33, 01.55)
};
\end{axis}
\end{tikzpicture}
\vspace{-0.4cm}
\caption{Instrument frequency in \dreal\ note annotations (grouped according to \texttt{MT3\_FULL\_PLUS} schema \cite{ymt3_plus}).}
\label{fig:inst_freq}
\end{figure*}
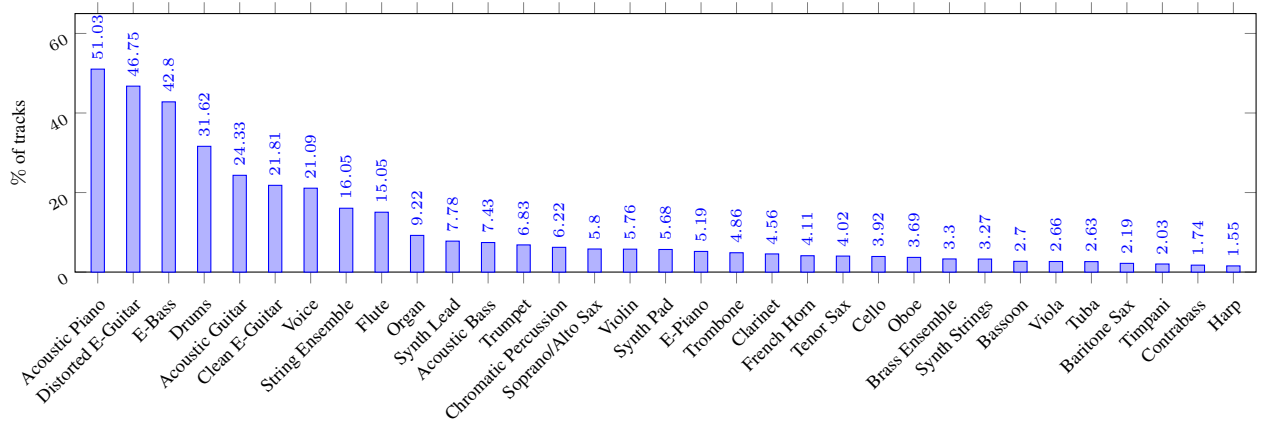

Our work differs from these architectural refinements by focusing on data scale and type, in particular on the relationship between synthetic and real training data. 
In related work, Maman and Bermano \cite{ben_transcription} worked towards a general-purpose transcription model, but at a smaller scale and without releasing final model weights. The use of synthetic data for (pre-)training has also been investigated in \cite{drum_transcription_synth_data, guitar_transcription_synth_data}, but for the single-instrument case. 
In addition, while most AMT models rely strictly on supervised teacher forcing, we introduce Reinforcement Learning via Group Relative Policy Optimization (GRPO) to align our model with high quality transcriptions, an unexplored technique in the music transcription literature.


\vspace{-0.15cm}
\section{Method}
\vspace{-0.1cm}
Similar to previous work, our model performs transcription by autoregressively predicting a MIDI-like token sequence given a mel-spectrogram of a short audio segment. Extending this framework, we investigate training on both large-scale synthetic and real music audio datasets (\refsection{ssec:datasets}). We further utilize reinforcement learning to improve results (\refsection{ssec:rl}) and allow optional conditioning on the instruments present in the transcription (\refsection{ssec:model}). 

\vspace{-0.15cm}
\subsection{Datasets}
\vspace{-0.1cm}
\label{ssec:datasets}
To train and evaluate our AMT models, we collect datasets with multi-instrument music audio and aligned note annotations covering a wide variety of musical genres.

\noindent\textbf{Synthetic data \dsynth:} 
As a primary contribution of this work, we investigate the impact of large-scale synthetic pre-training for AMT. To this end, we collect a dataset of roughly 1.45 million MIDI files (both from publicly available sources like Lakh MIDI \cite{lakh} and from commercial data providers) across different genres (with a focus on pop and Western classical music). In order to utilize this data for transcription, we develop an on-the-fly synthesis pipeline that randomly selects excerpts from MIDI files in our dataset, applies various augmentations on the symbolic level (pitch shifting, tempo changes, velocity adjustments, and instrument randomization), synthesizes the augmented MIDI using a randomly selected soundfont (from a collection of over 250 soundfonts), and finally applies a random detuning on the synthesized audio. As a consequence, each MIDI excerpt can be synthesized into an infinite number of potential audio realizations during training.

\noindent\textbf{Real music data \dreal:}
To complement our synthetic dataset \dsynth\ and to investigate the impact of different data types, we also utilize an internal dataset of 170\,000 real music audio recordings (totalling over 11\,000 hours) with aligned note annotations. Most of this data is obtained via audio-symbolic music synchronization. Here, we utilize a combination of linear interpolation between annotated bar line positions and (in the case of audio without bar annotations) dynamic time warping using chroma and onset features \cite{synctoolbox}. We used onset activation functions as in \cite{sync_onset_features}, which proved crucial for warping accuracy. We also filter out poor audio-symbolic pairs using a threshold on the time warping distance and by enforcing a maximum time dilation factor (i.e., after 8 seconds in one sequence, the alignment must have progressed 1 second in the other sequence). \reffigure{fig:inst_freq} shows statistics over the different instrument classes present in this dataset. Though the distribution is long-tailed, 17 different instrument groups are present in at least $5\%$ of tracks, indicating a dataset with diverse instrumentation.

\begin{figure}[t]
\centering
\begin{tikzpicture}
\begin{axis}[
    boxplot/draw direction=x,
    xlabel={Number of active instruments},
    ytick={1,2,3,4},
    yticklabels={\dtest, \drl, \dreal, \dsynth},
    ymin=0.5, ymax=4.5,
    xmin=0, xmax=17,
    height=4cm,
    width=\columnwidth,
    boxplot={box extend=0.4},
    tick label style={font=\small},
    label style={font=\small},
]
\addplot[boxplot prepared={
    draw position=4,
    lower whisker=1,
    lower quartile=1,
    median=4,
    upper quartile=6,
    upper whisker=15,
}, fill=blue!15, draw=blue!70] coordinates {};
\addplot[boxplot prepared={
    draw position=3,
    lower whisker=1,
    lower quartile=2,
    median=4,
    upper quartile=6,
    upper whisker=11,
}, fill=red!15, draw=red!70] coordinates {};
\addplot[boxplot prepared={
    draw position=2,
    lower whisker=2,
    lower quartile=5,
    median=7,
    upper quartile=9,
    upper whisker=12,
}, fill=teal!15, draw=teal!70] coordinates {};
\addplot[boxplot prepared={
    draw position=1,
    lower whisker=1,
    lower quartile=5,
    median=7,
    upper quartile=9,
    upper whisker=13,
}, fill=orange!15, draw=orange!70] coordinates {};
\addplot[only marks, mark=diamond*, mark size=1.5pt, fill=black]
    coordinates {(5.10,4) (4.37,3) (6.96,2) (7.03,1)};
\end{axis}
\end{tikzpicture}\vspace{-0.25cm}
\caption{Distribution of active instruments per track according to note annotations. Boxes span the interquartile range (p25--p75) with median line; whiskers extend to the 5th and 95th percentiles. Diamond markers indicate mean.}
\label{fig:instrument_count}
\end{figure}
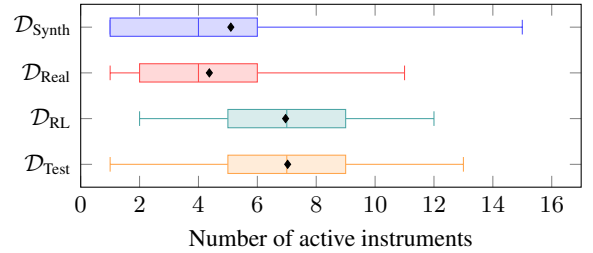

\noindent\textbf{Reinforcement learning data \drl:} 
We select 300 tracks from \dreal, manually verify that they have high annotation quality, and use these for reinforcement learning post-training, see \refsection{ssec:rl}.

\noindent\textbf{Test data \dtest:} Only few datasets for evaluating multi-instrument AMT exist and none cover the diversity of genres and instruments necessary in general-purpose transcription. We thus need to collect our own test data. Similar to \drl, we select 372 tracks from \dreal\ that have highly accurate annotations, subsequently remove them from \dreal, and use them as our test set. We additionally remove all tracks from our training sets that have similar titles as tracks in \dtest\ to ensure a clean split.

\reffigure{fig:instrument_count} shows statistics over the number of instruments in the tracks of our different datasets. We note that \dsynth\ contains a significant number of single-instrument tracks (which are synthesized as varying instruments during training due to our augmentation procedure). Our test set \dtest\ is focused on pieces featuring multiple active instruments, in line with real-world application scenarios.

\vspace{-0.15cm}
\subsection{Model}
\vspace{-0.1cm}
\label{ssec:model}
We employ a decoder-only Transformer architecture following standard practices, cf. \cite{musicgen}. We scale this architecture to four different sizes with 60M, 100M, 300M and 1.3B parameters, respectively (differing in the number of self-attention heads, stacked layers, and latent dimension). The model accepts two inputs: a mel-spectrogram representing a 5-second audio excerpt and a list of target instruments to be transcribed. To compute the spectrogram, the raw audio waveform (16kHz, mono) is converted into a mel-spectrogram using an STFT with $n_{\text{FFT}}=2048$, a hop size of 160 samples (yielding a 100Hz frame rate) and a mel filter-bank with 512 bins. For the instrument conditioning, we provide the model with the set of instruments present in the full track rather than the local excerpt. Consequently, for any given 5-second segment, the conditioning list may contain instruments that are not actively sounding in that specific segment. 

To feed the conditioning signals into the model, the mel-spectrogram is projected to the latent dimension of the transformer and then concatenated with embeddings for the present instrument classes (obtained via a learned lookup table). This embedding sequence is then used as prefix conditioning for the transformer model. 

To obtain training targets, we tokenize our note annotations with the usual MT3 tokenization scheme \cite{mt3}, but map the 128 MIDI instruments into 36 subgroups according to the \texttt{MT3\_FULL\_PLUS} taxonomy introduced in \cite{ymt3_plus} (same for the instrument conditioning).
The model is trained with teacher forcing over these tokens using a standard cross-entropy loss. We use 1M training steps with a batch size of 64, the AdamW optimizer with $\beta_1=0.9, \beta_2=0.95$, and a learning rate of $1e-4$ with a linear warmup of 2000 steps and a cosine schedule. Furthermore, each conditioning is independently dropped with a probability of 0.2. 

At inference time, we obtain transcriptions by feeding the mel-spectrogram and (optionally) instrument conditioning to the model and performing argmax decoding. The instrument conditioning allows both for customizing transcription results (e.g., transcribing only a subset of instruments) and for obtaining coherent transcriptions across segment boundaries, where instrument assignments predicted by the model might fluctuate. Moreover, we apply classifier-free-guidance \cite{cfg} to both conditions with a strength of $\acfg=2$. 

\vspace{-0.15cm}
\subsection{Evaluation Metrics}
\vspace{-0.1cm}
\label{sec:eval_metrics}
We evaluate our models using three standard instrument-agnostic metrics from the \texttt{mir\_eval} library \cite{mir_eval}. These metrics offer varying levels of stringency regarding temporal precision and note duration.

\noindent\textbf{Onset F1:} Following the evaluation protocol of MT3 \cite{mt3}, a predicted note is considered correct if its pitch matches the reference and its onset is within a $\pm 50$\,ms tolerance. 

\noindent\textbf{Offset F1:} This metric requires, in addition to the onset criterion, that the predicted offset falls within a tolerance of $\max(50\,\text{ms},\ 0.2 \times d_{\text{ref}})$, where $d_{\text{ref}}$ is the duration of the reference note. This rewards models that accurately capture note-off events and sustain.

\noindent\textbf{Frame F1:} To account for overall pitch activity, we compute frame-wise F1 at a resolution of $62.5$\,ms. Intuitively, this corresponds to the overlap between ground truth and predicted piano-rolls. This metric is more lenient than note-level metrics with regard to exact onset/offset times. 

\noindent\textbf{Drums Onset F1:} We separately report the onset F1 of predicted drum notes. Note that drum hits are onset-only.

\noindent\textbf{Multi F1:} An extension of the Offset F1 metric that additionally requires the instrument predicted for a note to be correct (for drum notes, only the onset is considered). This metric is the most representative of real-world transcription quality across a full multi-instrument mix.

Note that, because the tokenization scheme does not allow for multiple notes exciting the same pitch and instrument at the same time, we remove the shorter of multiple such overlapping notes from \dtest\ before computing these metrics. Thus, an optimal transcription model using the tokenization scheme could reach F1 scores of 1.0. \refsection{sec:results:overlapping_notes} investigates the impact of this choice.

\begin{figure*}[t]
    \centering
    \begin{tikzpicture}
        \definecolor{perfdarkblue}{rgb}{0.0,0.0,0.8}
        \definecolor{perfdarkred}{rgb}{0.8,0.0,0.0}
        \definecolor{perflightblue}{rgb}{0.7,0.85,1}
        \definecolor{perflightred}{rgb}{1,0.7,0.7}
        \pgfplotsset{%
            perf labels with/.style={%
                nodes near coords={\pgfmathprintnumber[fixed,precision=1]{\pgfplotspointmeta}},%
                every node near coord/.append style={%
                    font=\fontsize{5pt}{5.5pt}\selectfont, text=perfdarkblue, inner sep=0.1ex, yshift=0.7mm, anchor=south%
                }%
            },%
            perf labels without/.style={%
                nodes near coords={\pgfmathprintnumber[fixed,precision=1]{\pgfplotspointmeta}},%
                every node near coord/.append style={%
                    font=\fontsize{5pt}{5.5pt}\selectfont, text=perfdarkred, inner sep=0.1ex, yshift=-0.7mm, anchor=north%
                }%
            },%
            perf plot with/.style={%
                color=perflightblue, line width=0.95pt, mark=*, mark size=1.4pt,%
                mark options={draw=perfdarkblue, fill=perfdarkblue, line width=0.35pt},%
                perf labels with,%
            },%
            perf plot without/.style={%
                color=perflightred, line width=0.95pt, mark=square*, mark size=1.4pt,%
                mark options={draw=perfdarkred, fill=perfdarkred, line width=0.35pt},%
                perf labels without,%
            }%
        }%
        \begin{groupplot}[
            group style={
                group size=5 by 1,
                horizontal sep=0.4cm,
                y descriptions at=edge left
            },
            width={\dimexpr(1.45\textwidth-1.6cm-9mm)/5\relax},
            height=5cm,
            ymin=5, ymax=75,
            symbolic x coords={0\%, 1\%, 10\%, 50\%, 100\%},
            xtick=data,
            grid=major,
            grid style={dashed, gray!20},
            legend style={
                font=\scriptsize,
                legend columns=1,
                row sep=-1pt,
                inner sep=2pt
            },
            every axis/.append style={clip=false},
            every axis plot/.append style={mark size=1.4pt},
            tick label style={font=\tiny, rotate=30, anchor=east},
            label style={font=\scriptsize},
            title style={font=\scriptsize, yshift=-0.1cm},
            xlabel={}
        ]

        \nextgroupplot[title={Onset F1}, ylabel={F1 Score}]
        \addplot[perf plot with] coordinates {
            (0\%, 34.5)
            (1\%, 46.9)
            (10\%, 50.8)
            (50\%, 53.0)
            (100\%, 54.4)
        };
        \addplot[perf plot without] coordinates {
            (1\%, 21.2)
            (10\%, 45.4)
            (50\%, 52.1)
            (100\%, 53.2)
        };

        \nextgroupplot[
            title={Frame F1},
            legend to name=sharedlegend
        ]
        \addplot[perf plot with] coordinates {
            (0\%, 48.9)
            (1\%, 64.6)
            (10\%, 67.1)
            (50\%, 68.7)
            (100\%, 69.3)
        };
        \addlegendentry{With Pre-training on \dsynth}

        \addplot[perf plot without] coordinates {
            (1\%, 48.4)
            (10\%, 62.4)
            (50\%, 67.7)
            (100\%, 68.7)
        };
        \addlegendentry{Without Pre-training on \dsynth}

        \nextgroupplot[title={Offset F1}]
        \addplot[perf plot with] coordinates {
            (0\%, 16.1)
            (1\%, 33.4)
            (10\%, 37.9)
            (50\%, 41.0)
            (100\%, 42.3)
        };
        \addplot[perf plot without] coordinates {
            (1\%, 9.9)
            (10\%, 32.3)
            (50\%, 39.9)
            (100\%, 41.0)
        };

        \nextgroupplot[title={Drums F1}]
        \addplot[perf plot with] coordinates {
            (0\%, 21.0)
            (1\%, 35.4)
            (10\%, 41.2)
            (50\%, 42.6)
            (100\%, 43.3)
        };
        \addplot[perf plot without] coordinates {
            (1\%, 16.2)
            (10\%, 37.2)
            (50\%, 41.4)
            (100\%, 42.5)
        };

        \nextgroupplot[title={Multi F1}]
        \addplot[perf plot with] coordinates {
            (0\%, 16.2)
            (1\%, 32.8)
            (10\%, 37.7)
            (50\%, 40.4)
            (100\%, 41.6)
        };
        \addplot[perf plot without] coordinates {
            (1\%, 10.4)
            (10\%, 32.2)
            (50\%, 39.3)
            (100\%, 40.5)
        };

        \end{groupplot}

        \coordinate (perfplotsouth) at ($(group c1r1.outer south)!0.5!(group c5r1.outer south)$);
        \node[font=\scriptsize, anchor=north] (perf xlab) at (perfplotsouth) [yshift=-0.0cm] {\% of \dreal\ used};
        \node[anchor=north east, fill=white, fill opacity=0.85, text opacity=1, draw=black!40, line width=0.3pt, inner sep=2pt, rounded corners=1pt]
    at ($(group c5r1.north east) + (-0.1cm, -0.1cm)$)
    {\ref{sharedlegend}};
    \end{tikzpicture}\vspace{-0.3cm}
    \caption{Impact of pre-training on \dsynth\ across different sizes of the fine-tuning set \dreal\ ($\acfg=2$).}
    \label{fig:results:midi_pre_training}
\end{figure*}
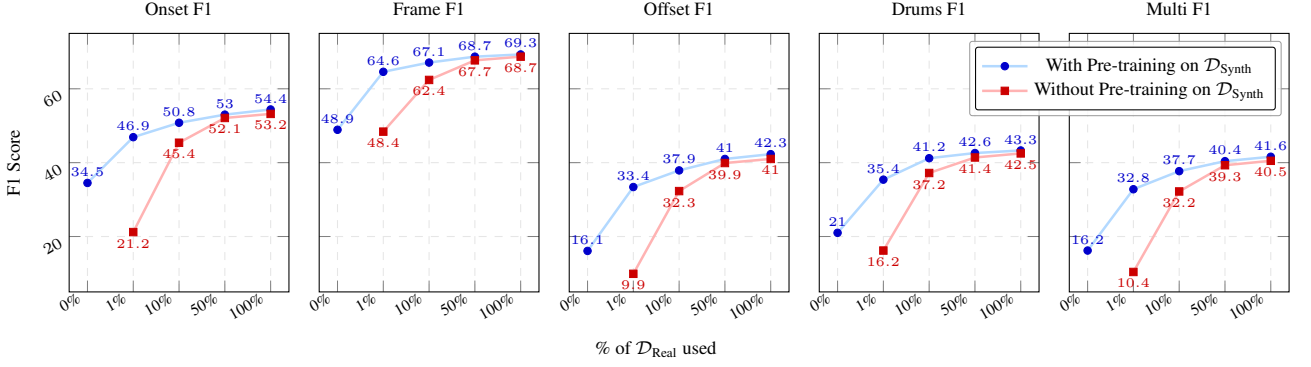

\vspace{-0.15cm}
\subsection{Reinforcement Learning Post-training}
\vspace{-0.1cm}
\label{ssec:rl}
We post-train our models on the separate high-quality dataset \drl\ using a policy gradient algorithm that combines
  the REINFORCE estimator~\cite{reinforce} with the group-relative advantage
  normalization introduced in GRPO~\cite{grpo}. At each training step, the model
  is set to evaluation mode and used to generate $G$ independent transcriptions
  for each audio segment in the batch by sampling from the autoregressive
  distribution with a fixed temperature $\tau$ ($\acfg=1$). For each generated sequence $y_{i,g}$ ($i \in \{1, ..., B\}$, $B$ being the batch size and $g \in \{1, \ldots, G\}$), a
  scalar reward is computed against the ground-truth MIDI reference as the sum of
  three note-level F-scores:
  \begin{equation}
      r_{i,g} = F_{\text{onset}} + F_{\text{frame}} + F_{\text{offset}}
  \end{equation}
  Following GRPO~\cite{grpo}, within-group advantages are computed by
  standardising the rewards across the $G$ outputs of the same segment
\begin{equation}
\begin{aligned}
    \hat{A}_{i,g} &= \frac{r_{i,g} - \mu_i}{\sigma_i + \varepsilon} \\
    \text{with } \mu_i = \tfrac{1}{G} \textstyle\sum_g r_{i,g}, \,\, &\sigma_i = \sqrt{\tfrac{1}{G} \textstyle\sum_g (r_{i,g} - \mu_i)^2}
\end{aligned}
\end{equation}  and $\varepsilon = 10^{-8}$. The model is then set back to training mode and
  updated via the standard REINFORCE~\cite{reinforce} objective, where the
  log-likelihood of each generated sequence is weighted by its advantage: 
  \begin{equation}
      \mathcal{L}_{\text{RL}} = \frac{1}{B\cdot G} \sum_{i,g}
      \hat{A}_{i,g} \cdot \mathcal{L}_{\text{CE}}(\theta,\tau; y_{i,g}),
  \end{equation}
  where $\mathcal{L}_{\text{CE}}(\theta, \tau; y) = \sum_{t=1}^{T}\frac{\exp(l_{\theta}(y^t)/\tau)}{\sum_{y^{'} \in \mathcal{V}}\exp(l_\theta(y^{'})/\tau)}$ is the average
  cross-entropy of the model with temperature $\tau$ over the generated sequence $(y^1, ..., y^T)$, $\mathcal{V}$ is the vocabulary of tokens and $l_\theta(k)$ is the logit of the k-th element of the vocabulary. 
  Unlike full GRPO, no importance-sampling ratio clipping or KL divergence
  penalty against a frozen reference policy is applied. In practice we perform this optimization with $G=8$ over segments of 5 seconds with a temperature $\tau=0.75$ and a batch size of 8. 

\vspace{-0.15cm}
\section{Experimental Results}
\vspace{-0.1cm}

In our experiments, we generally pre-train our model on \dsynth, fine-tune on \dreal, post-train on \drl, and evaluate on \dtest. Unless stated otherwise, all results are for the 1.3B parameters model with instrument conditioning.

\vspace{-0.15cm}
\subsection{Main Results}
\vspace{-0.1cm}

\begin{table}[t]
\centering
\caption{Comparison of results on \dtest\ for MuScriptor (1.3B) across different training stages and CFG settings.}
\label{tab:results:main}
\resizebox{\columnwidth}{!}{%
\begin{tabular}{@{}lccccccc@{}}
\toprule
\multirow{2}{*}{\textbf{Model}} & \multirow{2}{*}{\textbf{$\acfg$}} & \multicolumn{5}{c}{\textbf{F1 Score ($\uparrow$)}} \\ \cmidrule(l){3-7} 
 &  & \textbf{Onset} & \textbf{Frame} & \textbf{Offset} & \textbf{Drums} & \textbf{Multi} \\ \midrule
YourMT3+ {\tiny (YPTF.MoE+Multi (noPS))} & -- & 32.52 & 45.54 & 17.79 & 41.4 & 21.9 \\ \midrule
MuScriptor trained on &  &  &  &  &  & \\
\multirow{2}{*}{\hspace{1em}\dsynth} & 1 & 26.1 & 51.3 & 14.2 & 23.1 & 15.2 \\
 & 2 & 34.5 & 48.9 & 16.1 & 21.0 & 16.2 \\ \addlinespace 
\multirow{2}{*}{\hspace{1em}\dsynth{} + \dreal} & 1 & 52.5 & 69.4 & 42.0 & 44.7 & 41.7 \\ 
 & 2 & 54.4 & 69.3 & 42.3 & 43.3 & 41.6 \\ \addlinespace 
\multirow{2}{*}{\hspace{1em}\dsynth{} + \dreal{} + \drl} & 1 & \textbf{60.4} & \textbf{73.3} & \textbf{49.0} &  \textbf{50.2} & \textbf{48.2}     \\ 
 & 2 & \textbf{60.4} & 72.4 & 48.6  & 49.6 & 47.8 \\ \bottomrule
\end{tabular}%
}
\end{table}

Our main results are summarized in \reftable{tab:results:main}. We observe that each training stage improves results over the previous one. In particular, while the model trained exclusively on synthetic data is already competitive with the state-of-the-art AMT baseline from \cite{ymt3_plus} in terms of frame F1 score, we note that fine-tuning on \dreal\ is essential for performance and improves all metrics by roughly 20 percentage points. In addition, the reinforcement learning phase on the high-quality dataset \drl\ further improves results, leading to our best overall model. The effects of the various training stages are also illustrated in \reffigure{fig:pianoroll_song20}. Qualitatively, we observe that reinforcement post-training reduces false negatives and improves the onset precision of our model.

\reftable{tab:results:main} also illustrates the impact of applying classifier-free-guidance at inference time. Comparing $\acfg=1$ (i.e., CFG off) with $\acfg=2$, we observe improved metrics for most model configurations. In particular, for the model trained on \dsynth\ only, the onset F1 score jumps from $26.1$ to $34.5$. For the RL model, there are no further improvements through CFG, meaning that one can avoid the second forward pass at inference time here.

\vspace{-0.15cm}
\subsubsection{Analysis of Synthetic Pre-training}
\vspace{-0.1cm}

\reffigure{fig:results:midi_pre_training} gives a detailed view of the impact of pre-training on synthetic data. Here we compare models trained with different amounts of real music audio data (x-axis) that are either initialized from scratch or pre-trained on \dsynth. We observe that pre-training on synthetic data is highly effective, especially when only little real music audio data is available (e.g., improving offset F1 from $9.9$ to $33.4$ when only $1\%$ of \dreal\ is used). However, we also see that training on synthetic data alone is not sufficient and that results steadily improve the more real music data is used. Nevertheless, the best results are obtained when combining both data sources (e.g., at $100\%$ of \dreal\ used, the pre-trained model achieves $42.3$ offset F1 compared to $41$ without pre-training).

\vspace{-0.15cm}
\subsubsection{Benchmark Datasets}
\vspace{-0.1cm}

\begin{table}[t]
\centering
\caption{Per-dataset comparison of YourMT3+ (YPTF.MoE+Multi (noPS)) and MuScriptor (1.3B, trained on \dsynth{} + \dreal{} + \drl{}, $\acfg=1$).}
\label{tab:results:per_dataset}
\resizebox{\columnwidth}{!}{%
\begin{tabular}{@{}lccccc@{}}
\toprule
 & \multicolumn{5}{c}{\textbf{F1 Score ($\uparrow$)}} \\ \cmidrule(l){2-6}
\textbf{Dataset / Model} & \textbf{Onset} & \textbf{Frame} & \textbf{Offset} & \textbf{Drums} & \textbf{Multi} \\ \midrule
Bach10 \cite{bach10} \\
\hspace{1em}YourMT3+    & \textbf{59.8} & 66.0          & \textbf{48.0} & --            & 26.4 \\
\hspace{1em}MuScriptor  & 43.1          & \textbf{85.0} & 36.0          & --            & \textbf{34.7} \\ 
Dagstuhl ChoirSet \cite{dagstuhl_choirset} \\
\hspace{1em}YourMT3+    & \textbf{22.3} & 51.0          & 10.8          & --            & 2.6  \\
\hspace{1em}MuScriptor  & 14.4          & \textbf{80.7} & \textbf{11.5} & --            & \textbf{11.5} \\ 
PHENICX-Anechoic \cite{phenicx_anechoic} \\
\hspace{1em}YourMT3+    & \textbf{56.7} & 58.9          & 18.7          & --            & 12.2 \\
\hspace{1em}MuScriptor  & 56.1          & \textbf{74.6} & \textbf{32.6} & --            & \textbf{25.7} \\ 
RWC-P \cite{rwc_revisited} \\
\hspace{1em}YourMT3+    & 36.1          & 51.6          & 20.6          & 36.5          & 19.1 \\
\hspace{1em}MuScriptor  & \textbf{46.1} & \textbf{61.2} & \textbf{25.6} & \textbf{42.1} & \textbf{25.6} \\ 
RWC-C \cite{rwc_revisited} \\
\hspace{1em}YourMT3+    & \textbf{71.7} & \textbf{71.3} & \textbf{44.3} & 8.7           & \textbf{40.5} \\
\hspace{1em}MuScriptor  & 67.7          & 70.5          & 36.9          & \textbf{23.7} & 36.0 \\ 
RWC-G \cite{rwc_revisited} \\
\hspace{1em}YourMT3+    & 36.9          & 49.4          & 20.5          & 25.6          & 17.2 \\
\hspace{1em}MuScriptor  & \textbf{44.7} & \textbf{58.7} & \textbf{24.4} & \textbf{29.2} & \textbf{23.7} \\ 
RWC-J \cite{rwc_revisited} \\
\hspace{1em}YourMT3+    & 52.9          & 57.2          & 31.1          & 30.6          & 26.4 \\
\hspace{1em}MuScriptor  & \textbf{59.4} & \textbf{62.7} & \textbf{33.9} & \textbf{31.3} & \textbf{31.8} \\ 
RWC-R \cite{rwc_revisited} \\
\hspace{1em}YourMT3+    & 46.1          & 61.5          & \textbf{28.6} & \textbf{51.3} & \textbf{23.1} \\
\hspace{1em}MuScriptor  & \textbf{47.1} & \textbf{68.6} & 24.8          & 36.5          & 20.3 \\ \bottomrule
\end{tabular}%
}
\end{table}

As explained in \refsection{ssec:datasets}, there is a lack of existing benchmark datasets for AMT covering a multitude of genres and instrumentation. We therefore compare our models on \dtest\ for most purposes. Nevertheless, we compare against several existing datasets with music audio and note annotations in \reftable{tab:results:per_dataset}. We purposefully choose datasets that are not contained in either model's training set to check cross-domain generalization.
We observe substantial gains, especially in terms of frame and multi F1 scores across several datasets. For example, we improve frame F1 on Dagstuhl ChoirSet from $51.0$ with YourMT3+ to $80.7$. However, onset and offset scores remain lower, pointing to the difficulty of annotating precise note on- and offsets for certain styles of music (such as chorals, cf.~\cite{dagstuhl_choirset}).

\vspace{-0.15cm}
\subsubsection{Instrument Conditioning}
\vspace{-0.1cm}

\begin{table}[t]
\centering
\caption{Impact of instrument conditioning on a MuScriptor model trained exclusively on \dreal, $\acfg=2$.}
\label{tab:results:instrument_conditioning}
\resizebox{\columnwidth}{!}{%
\begin{tabular}{@{}lccccc@{}}
\toprule
\multirow{2}{*}{\textbf{Instrument Conditioning}} & \multicolumn{5}{c}{\textbf{F1 Score ($\uparrow$)}} \\ \cmidrule(l){2-6} 
 &  \textbf{Onset} & \textbf{Frame} & \textbf{Offset} & \textbf{Drums} & \textbf{Multi} \\ \midrule
Off                                 & 51.6 & 66.5 & 40.1 & 40.6 & 38.7 \\ \addlinespace 
On                                 & 53.2 & 68.7 & 41.0 & 42.5 & 40.5 \\ \bottomrule 
\end{tabular}%
}
\end{table}

In \reftable{tab:results:instrument_conditioning} we compare results with and without supplying instrument conditioning to our trained model at inference time. Note that the model is trained to work in both settings via conditioning dropout during training. Here, for the results with conditioning, we supply the actual instruments present in the track as given in the ground truth annotations, corresponding to practical usecases where the instruments present in a track are often known from metadata. As expected, supplying instrument conditioning improves quantitative scores.

\vspace{-0.15cm}
\subsection{Ablation Study}
\vspace{-0.1cm}
\subsubsection{Model Size}
\vspace{-0.1cm}

\begin{table}[t]
\centering
\caption{Performance comparison across different model scales, trained exclusively on \dreal, $\acfg=2$}
\label{tab:results:model_scale}
\resizebox{\columnwidth}{!}{%
\begin{tabular}{@{}lccccc@{}}
\toprule
\multirow{2}{*}{\textbf{Model}} & \multicolumn{5}{c}{\textbf{F1 Score ($\uparrow$)}} \\ \cmidrule(l){2-6} 
 & \textbf{Onset} & \textbf{Frame} & \textbf{Offset} & \textbf{Drums} & \textbf{Multi} \\ \midrule
MuScriptor (60M)                       & 47.7 & 65.7 & 35.3 & 39.8 & 35.2 \\ \addlinespace 
MuScriptor (100M)                      & 51.2 & 67.2 & 38.7 & 41.5 & 38.2  \\ \addlinespace 
MuScriptor (300M)                      & 52.4 & 68.0 & 40.3  & 42.0 & 39.7 \\ \addlinespace 
MuScriptor (1.3B)                      & 53.2 & 68.7 & 41.0 & 42.5 & 40.5 \\ \bottomrule 
\end{tabular}%
}
\end{table}

\reftable{tab:results:model_scale} shows the impact of scaling our model architecture towards different sizes. As expected, our transformer model yields better results at larger model scales. Nevertheless, even the smallest model with 60M parameters provides strong performance with a frame F1 score of $65$. We thus expect our smaller MuScriptor model to also be applicable in lower-resource settings.

\vspace{-0.15cm}
\subsubsection{Audio Representation}
\vspace{-0.1cm}

\begin{table}[t]
\centering
\caption{Performance comparison for different audio input representations. MuScriptor (300M) models trained exclusively on \dreal, $\acfg=2$.}
\label{tab:results:audio_representation}
\resizebox{\columnwidth}{!}{%
\begin{tabular}{@{}lccccc@{}}
\toprule
\multirow{2}{*}{\textbf{Input}} & \multicolumn{5}{c}{\textbf{F1 Score ($\uparrow$)}} \\ \cmidrule(l){2-6} 
  & \textbf{Onset} & \textbf{Frame} & \textbf{Offset} & \textbf{Drums} & \textbf{Multi} \\ \midrule
Mel-spectrogram (default) & 52.4 & 68.0 & 40.3 & 42.0 & 39.7 \\ \addlinespace
Magnitude CQT & 51.7 & 67.5 & 39.8 & 37.2 & 38.2 \\ \addlinespace
Encodec & 39.7 & 58.0 & 27.5 & 37.9 & 28.9 \\ \addlinespace
MERT & 48.5 & 63.9 & 36.4 & 43.1 & 36.8 \\ \bottomrule
\end{tabular}%
}
\end{table}

In line with \cite{mt3, ymt3_plus}, we use mel-scaled spectrograms as input to our transcription model. \reftable{tab:results:audio_representation} illustrates the impact of this choice. Though conceptually similar, a magnitude CQT with the same frame rate of 100\,Hz and three bins per octave (252 bins in total) performed slightly worse in our experiments. Alternative setups with a neural audio codec (Encodec \cite{encodec}, with a frame rate of 50\,Hz) or music embeddings (MERT \cite{mert}, with a frame rate of 75\,Hz) as input yielded even worse results. This suggests that a representation that is close to actual signal properties might work best for the transcription task.

\vspace{-0.15cm}
\subsubsection{Overlapping Notes}
\label{sec:results:overlapping_notes}
\vspace{-0.1cm}

As explained in \refsection{sec:eval_metrics}, we drop short notes from \dtest\ that overlap with another note of the same instrument and pitch. If we keep these notes, evaluation metrics drop. For example, for our 1.3B model trained on all datasets (cf. \reftable{tab:results:main}, $\acfg=1$) onset F1 drops from $60.4$ to $51.8$, offset F1 from $49.0$ to $41.9$ and multi F1 from $48.2$ to $42.0$. This demonstrates that overlapping notes for the same instrument regularly occur in practice. A different tokenization scheme is needed to enable transcription of such notes.

\vspace{-0.15cm}
\section{Conclusion}
\vspace{-0.1cm}

In this work, we presented MuScriptor, an open weight model for general-purpose, multi-instrument music transcription. To train MuScriptor, we collected synthetic and real music audio datasets with aligned note annotations, and used a reinforcement learning approach for post-training refinement. Our model outperforms a state-of-the-art baseline by a wide margin and is configurable at inference time through optional instrument conditioning. Moreover, we provided an analysis of the impact of synthetic music data for pre-training, as well as several ablations.

To our knowledge, MuScriptor is the first effective open weight model for music transcription across various genres. We hope that MuScriptor will enable various applications for both researchers in music information retrieval and music practitioners. Future work may explore extending the instrument vocabulary, allowing concurrent note activity for a single instrument class, and extending the segment size to enable faster inference.

\bibliography{ISMIRtemplate}

\end{document}